# PHOTON SPLITTING IN SOFT GAMMA REPEATERS

MATTHEW G. BARING* and ALICE K. HARDING
*NASA Goddard Space Flight Center, Code 665,
Greenbelt, MD 20771, U.S.A.*

**Abstract.** The exotic quantum process of photon splitting $\gamma \to \gamma\gamma$ has great potential to explain the softness of emission in soft gamma repeaters (SGRs) if they originate in neutron stars with surface fields above the quantum critical field $B_{\rm cr} = 4.413 \times 10^{13}$ Gauss. Splitting becomes prolific at such field strengths: its principal effect is to degrade photon energies, initiating a cascade that softens gamma-ray spectra. Uniform field cascade calculations have demonstrated that emission could be softened to the observed SGR energies for fields exceeding about $10^{14}$ Gauss. Recently, we have determined splitting attenuation lengths and maximum energies for photon escape in neutron star environments including the effects of magnetospheric dipole field geometry. Such escape energies $\varepsilon_{\rm esc}$ suitably approximate the peak energy of the emergent spectrum, and in this paper we present results for $\varepsilon_{\rm esc}$ as a function of photon emission angles for polar cap and equatorial emission regions. The escape energy is extremely insensitive to viewing perspective for equatorial emission, arguing in favour of such a site for the origin of SGR activity.

## 1. Introduction

Soft gamma repeaters (SGRs), have recently become extremely topical in the high energy astrophysics community, largely due to the identification of multiple counterparts to these gamma-ray transients in other wavebands. These associations include radio supernova remnants (e.g. G10.0-0.3 for SGR1806-20, see Kulkarni and Frail, 1993) and a compact X-ray region within the X-ray/radio map of the LMC remnant N49 for the 5th March 1979 repeater (Rothschild *et al.*, 1994). The cumulative effect of these and other observations has been a strengthening of the idea that SGRs are of neutron star origin. If such an association is admitted and the eight second periodicity of the soft gamma-ray tail to the Mar. 5, 1979 repeater is assumed to result from magnetic dipole radiation spin-down throughout the life of its parent neutron star, then a field estimate (Duncan and Thomson, 1992) of $B \sim 6 \times 10^{14}$ Gauss is obtained using the estimated age of N49 of 5400 years. Such fields, exceed the quantum critical field strength $B_{\rm cr} = m^2 c^3 / e\hbar = 4.413 \times 10^{13}$ Gauss, and dwarf those found in most radio pulsars.

In such conditions, the exotic process of magnetic photon splitting $\gamma \to \gamma\gamma$ acts effectively (Baring, 1991) below the threshold of single photon pair production $\gamma \to e^+ e^-$ around 1 MeV, attenuating gamma-rays and degrading them to soft gamma-ray energies while at the same time polarizing the emission. Recently, Baring (1995) outlined the thermostatic action of splitting as a possible reason for why SGR emission is so soft: $\gamma \to \gamma\gamma$ is unavoidable in supercritical fields, and the emergent spectra are then largely insensitive to the field strength above $\sim 5 B_{\rm cr}$. Baring (1995) observed that a field estimate of $B \sim 4 B_{\rm cr}$ would be required to roughly fit the ICE observations (Fenimore *et al.*, 1994) of SGR1806-20. As a further exploration of these uniform field calculations, Baring and Harding (1995) calculated splitting attenuation lengths for photons as a function of their emission point and propagation angle near the neutron star surface, including the effect of dipole field geometry. An aspect of this analysis is presented in this paper, namely the computation of the maximum energy for photon escape $\varepsilon_{\rm esc}$ from neutron star magnetospheres (below which the SGR emission will generally appear), as a function of photon emission angles for polar cap and equatorial emission regions. Our analysis strongly suggests that SGR emission emanates from equatorial regions if it is the result of splitting cascades.

* Compton Fellow, USRA;  Email: *Baring@lheavx.gsfc.nasa.gov*

## 2. Photon Splitting and Escape Energies

Photon splitting is forbidden in field-free regions, but has a substantial non-zero rate and is a powerful polarizing mechanism in a magnetized vacuum. For $\varepsilon B \lesssim B_{\rm cr} mc^2$ and $B \lesssim B_{\rm cr}$, where $\varepsilon$ is the incident photon energy, the splitting attenuation coefficients (i.e. rates divided by $c$) for different photon polarization modes assume simple forms (using rates from Adler 1971, Baring 1991): $T_{sp} \propto \varepsilon^5 (B/B_{\rm cr})^6 \sin^6 \theta_{\rm kB}$, where $\theta_{\rm kB}$ is the angle between the photon momentum and the magnetic field vectors. Reducing $\theta_{\rm kB}$ or $B$ dramatically increases the photon energy required for splitting to operate; such sensitivities suggest that considerations of magnetospheric geometry are important for analyses of splitting, motivating the work presented here. High field ($B \gtrsim B_{\rm cr}$) corrections (e.g. see Adler 1971, Baring 1995) to the above formula for splitting diminish its dependence on $B$.

The simplest quantity that defines the effectiveness of photon splitting in SGRs from neutron star magnetospheres is its attenuation length $L$, defined to be the path length over which the optical depth of $\gamma \to \gamma\gamma$ is unity. We consider photons that originate on the neutron star surface, and propagate (generally non-radially) outward. Attenuation lengths are explored in detail by Baring and Harding (1995), where $L$ was generally found to be shorter for polar (as opposed to equatorial) emission where the surface field strength is higher. At low energies the splitting rate is low, and photons can escape the magnetosphere. For each set of initial conditions, i.e. photon location and propagation direction, there is a critical energy $\varepsilon_{\rm esc}$, called the *escape energy*, below which the optical depth to infinity is always less than unity and photons escape the magnetosphere; this occurs because of the decay of the dipole field away from the stellar surface.

Figure 1 depicts escape energies for polar ($\theta = 0°$) and equatorial emission ($\theta = 90°$) as a function of the initial propagation angle $\theta_{\rm k}$ to the dipole axis, where $90° - \theta$ is the latitude of emission. The neutron star radius was set at $10^6$ cm. The plots discriminate between the two polarization states of the photons, namely $\perp$ or $\parallel$, where the photon's electric field vector is respectively parallel or orthogonal to $\mathbf{k} \times \mathbf{B}$. In the polar case, $\varepsilon_{\rm esc}$ clearly declines with both $\theta_k$ and $B$. In the case of equatorial emission, the results in Fig. 1 indicate a remarkable insensitivity to $\theta_k$. This is largely due to the fact that even if photons start out moving along the field, they soon propagate obliquely to the field because of its curvature and so cannot escape unless they are below the $\theta_k = 90°$ escape energy. The curves in Fig. 1 indicate that for low $B$, photons with $\perp$ polarization are more likely to split than those in the $\parallel$ state, while for $B \gg B_{\rm cr}$ this situation is reversed. Also note that for $\theta_k \sim 90°$, the escape energies for polar and equatorial emission are similar when $B \gg B_{\rm cr}$; this arises because then the splitting of photons above $\varepsilon_{\rm esc}$ occurs in regions away from the stellar surface.

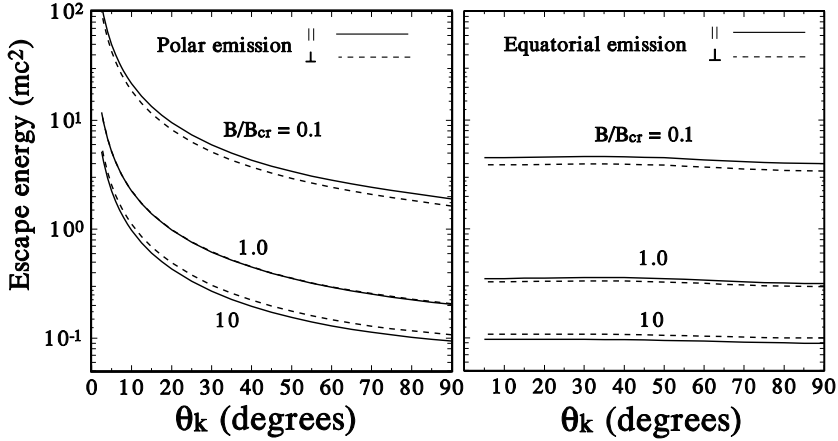

Fig. 1. The energy, below which photons escape from the magnetosphere without splitting, for photons of observable polarization ($\perp$ and $\parallel$ are defined in the text) originating at the pole (left) and the equator (right), as functions of the polar angle $\theta_k$ of the photon. The curves tend to merge above $10 B_{\rm cr}$ due to the saturation of the splitting rate at high fields, and therefore are not displayed. The curves for polar emission diverge near $\theta_k = 0°$ because the photons are almost parallel to the field lines throughout their path.

## 3. Discussion

Our calculations of the photon splitting escape energy in a neutron star dipole field give a realistic estimate of the degree of spectral softening achievable through a photon splitting cascade. They go beyond the work of Baring (1995), who assumed a homogeneous field of finite extent $2 \times 10^6$ cm and found that the escaping photon spectrum is quasi-thermal and peaks roughly at the escape energy. Our escape energies for polar emission at $\theta_k = 90°$ are $\sim 2.5$ times those of Baring, primarily because a dipole field decreases signif-icantly over the region that he assumed to be homogeneous. The surface field strength required to fit observed SGR spectra with photon splitting cascade spectra will therefore be somewhat higher than the $B \sim 4\,B_{\rm cr}$ obtained by Baring in fitting spectra from SGR 1806-20.

Fig. 1 exhibits a strong variation in escape energies for photons emitted at different angles near the magnetic pole of the neutron star, but almost *no* variation for photons emitted at the dipole equator. This result is purely a function of the geometry of the dipole field curvature: near the pole, the field lines are diverging rapidly and different emission directions sample very different field orientations, whereas at the equator, the field looks nearly the same in all directions. This behaviour has important implications for splitting cascade models of SGR spectra. It predicts that the cascade spectra from emission near the equator will be insensitive to observer angle, yielding the same peak energy in all directions. Thus, the spectra would not vary from burst to burst, even if the neutron star orientation changes (i.e. the star rotates). This is consistent with what is observed in different SGR bursts (e.g. Fenimore, Laros and Ulmer, 1994; Kouveliotou *et al.*, 1994 for SGR 1806-20), and also with phase-resolved spectroscopy of the periodic soft gamma-ray tail of the Mar. 5, 1979 outburst (Mazets *et al.*, 1982). Hence the principal conclusion of this paper is that emission near the dipole equator would be strongly favoured for photon splitting cascade models of SGRs.

The initial angular distribution of the photons, which depends on the emission mechanism operating, will play a crucial role in determining the escape energy. For example, resonant Compton upscattering (e.g. Dermer, 1990) produces $\gamma$-ray photons from soft thermal photons above a neutron star surface that are beamed at small angles to the field. Cyclotron radiation from relativistic electrons in low Landau levels has similar beaming. Escape energies for these photons will be much higher at the pole than at the equator (see Fig. 1), where even photons propagating initially parallel to the field have low $\varepsilon_{\rm esc}$. Consideration of such angular distributions forms the next stage of our program of research into splitting cascades and SGR emission.

## References

Adler, S. L.: 1971, *Ann. Phys.* **67**, 599.
Baring, M. G.: 1991, *Astron. & Astrophys.* **249**, 581.
Baring, M. G.: 1995, *Astrophys. J. (Lett.)* **440**, L69.
Baring, M. G. and Harding, A. K.: 1995, to appear in *"High Velocity Neutron Stars and Gamma-Ray Bursts"* eds. Rothschild, R. et al., AIP, New York.
Dermer, C. D.: 1990, *Astrophys. J.* **360**, 197.
Duncan, R. C. and Thompson, C.: 1992, *Astrophys. J. (Lett.)* **392**, L9.
Fenimore, E. E., Laros, J. G. and Ulmer, A.: 1994, *Astrophys. J.* **432**, 742.
Kouveliotou, C., et al.: 1994, *Nature* **368**, 125.
Kulkarni, S. and Frail, D.: 1993, *Nature* **365**, 33.
Mazets, E. P. et al.: 1982, *Astrophys. Space Sci.* **84**, 173.
Rothschild, R. E., et al.: 1994, *Nature* **368**, 432.